# Customer Validation, Feedback and Collaboration in Large-Scale Continuous Software Development

David Molamphy
Dell Technologies & University of Limerick
Limerick, Ireland
david.molamphy@dell.com

Advisor:
Prof. Brian Fitzgerald
LERO, University of Limerick
Limerick, Ireland
brian.fitzgerald@ul.ie

Advisor:
Prof. Kieran Conboy
LERO, University of Galway
Galway, Ireland
kieran.conboy@universityofgalway.ie

*Abstract*—The importance of continuously incorporating customer feedback in the software development process is well established and firmly grounded in concepts such as agile and DevOps. In large-scale organizations such as Dell Technologies however, an array of challenges remain unsolved relating to this crucial aspect of software development. Despite a wide variety of tools and techniques available for collecting and analyzing customer feedback, in large-scale organizations implementing agile and continuous software development practices, harmful disconnects, discrepancies and processes exist. Such challenges negatively impact on an organizations ability to regularly deploy incremental improvements to their software products which meet customer needs. In this Professional Doctorate research program, wherein the researcher is a practitioner within Dell Technologies, we explore the challenges of continuously integrating customer feedback in a large-scale global organization with over 100,000 employees and hundreds of software products. Leveraging an Action Research approach, we will propose a model to enhance the continuous incorporation of customer feedback and validation, providing organizations with the ability to frequently deliver incremental software improvements which satisfy the needs of its customers, measurable by metrics such as customer satisfaction, product adoption, bugs/defect escapes, production incidents and deployment frequency.

## I. INTRODUCTION & PROBLEM STATEMENT

The agile manifesto emphasizes rapid, high-quality software delivery, which is flexible to customer needs and market changes [1]. Agile methodologies focus on short development cycles, frequent deliveries, cross-team collaboration, and constant customer interaction to ensure satisfaction [2]. Despite the appeal of this customer-centric approach, implementing agile practices and cultural transformations pose many challenges.

Agile methods, originally developed for smaller organizations, are now adopted by companies of all sizes [3]. Large-scale agile implementation is complex, with challenges such as the assumption of small, co-located teams [4] and collaboration issues among other teams and business units [5]. To address these issues, many organizations adopt frameworks like Scrum-at-Scale, LeSS, and SAFe, or custom implementations of such methods, with reportedly varying degrees of success [6].

Continuous software development (CSD), though open to interpretation [7], encompasses agile, lean, and DevOps practices [8]. It promotes rapid software delivery and the incorporation of user feedback to improve the product [9]. BizDev extends this concept, focusing on integrating and aligning broader organizational roles into the development process [7].

Incorporating customer feedback is crucial for software product development, aligning with Drucker's principle that customers define the success of a business [10]. Concepts such as DevOps, which emphasizes a collaborative culture across the software development organization, supports the aforementioned agile principles [11]. The benefits of customer feedback include reduced development waste, improved product adoption, and increased customer satisfaction [1, 12, 13]. However, effectively integrating feedback techniques and tools into large-scale continuous software development remains a challenge [14].

Various techniques which pre-date the agile and continuous movement exist in software development for gathering customer feedback, such as surveys, interviews, and usage data [15]. User Acceptance Testing (UAT) is a traditional method but faces scaling issues and depends on customer participation, which is often limited [16]. It is an inherently manual and gated process, and large-scale organizations often struggle with enforcing customer involvement [17].

Further to the BizDev concept [9], large-scale organizations may encounter strategic and structural issues which impact the ability of software teams to embrace opportunities for customer involvement. Recognition of the need for disparate business units to adopt a shared strategy and align to the continuous and agile practices necessary to facilitate continuous customer validation presents a challenge in large enterprises [5]. Structural issues may pose additional challenges, both in terms of sheer scale and global distribution in large organizations, but also the potential levels software teams are removed from customers. In addition, relationship management between product development and customer organizations is a known challenge [6], presenting a possible discontinuity in the agile and continuous concept.

A key concept of continuous software development is continuous deployment, providing the ability to rapidly and automatically deploy incremental code changes to production [18]. Often however, opportunities to establish the validity of a software products design and functionality from a customer perspective is only established once the change has been deployed to production [19]. This presents a challenge not only by creating less-than-optimal conditions for the prioritization and grooming of features, but also introduces the risk of dissatisfied customers having to request fixes or enhancements following a deployment. Techniques such as A/B testing are leveraged by some organizations in order to understand the relationship between a change and its use by customers, effectively as a form of hypothesis testing [20]. Other approaches such as continuous delivery - constantly maintaining a product code base in a deployable state [21], provides opportunities for 'non-production' validation techniques, such as demo and sandbox environments which can be leveraged as part of UAT processes. An important consideration which requires further exploration, is the application of techniques such as UAT in continuous software development approaches, as traditional UAT approaches requires a waterfall approach to its implementation and management, in addition to a high degree of customer/user participation and manual effort [16].

Despite the adoption of agile and DevOps methodologies to enhance software delivery, incorporating customer feedback remains problematic [19]. Challenges include maintaining user engagement, resource allocation, regulatory compliance, selecting appropriate metrics and data literacy [14]. In large-scale continuous software development, additional challenges are likely to emerge. Given this, we believe significant opportunities exist to improve the approaches and techniques used to collect customer feedback and determine customer validity of software products in the continuous software development process.

## II. RESEARCH HYPOTHESIS

We hypothesize that development of a model which addresses the challenges of continuously integrating customer feedback in large-scale continuous software development practices will lead to improvements in the frequency and quality of software delivery, measurable through bugs/defect escapes, development velocity, customer satisfaction and other key indicators (to be defined in the exploratory research phase).

*Null Hypothesis (H0):* Implementation of a continuous feedback and validation model does not result in significant improvements in key software development indicators.

*Alternative Hypothesis (H1):* Implementation of a continuous feedback and validation model results in significant improvements in key software development indicators.

## III. RESEARCH QUESTIONS

RQ1: What customer feedback and validation techniques are currently employed by software teams in a large-scale organization practicing continuous software development?

RQ2: What are the primary challenges faced in the organization in continuously incorporating customer feedback, and what opportunities exist to enhance these processes?

RQ3: What are the critical components of a customer-centric model for continuous validation in a large-scale, continuous software development organization, and what are the possible key indicators to measure the impact of such a model?

RQ4: What is the impact of implementing a model for continuous customer validation in a large-scale continuous software development organization?

## IV. EXPECTED CONTRIBUTIONS

The expected contributions of this research project to both literature and practice are as follows:

### A. Identification of customer validation challenges and opportunities in a large-scale development context

Little research exists which investigates the techniques and approaches to customer validation and feedback in large-scale software development, particularly so in the continuous software development domain. The intent of the research project is to identify the challenges and opportunities with regard to continuous customer feedback and validation in a large global software development organization.

### B. Identification of customer validation techniques at scale

By way of examining the practices of a wide range of software teams across the organization coupled with quantitative data such as indicators of product success and customer sentiment, we will identify the techniques used from a customer validation perspective, and examine the relationship between these techniques and the associated outcomes.

### C. Development of a customer-centric model for continuous validation in a large-scale development context

Building on the findings of the primary data collection and complemented by a comprehensive literature review, we intend to develop a model for customer-centric continuous feedback and validation. The model will provide practitioners with a set of techniques and technical approaches for continuous data collection and collaboration with customers, and may leverage a variety of tools and systems to provide teams with the ability to automate aspects of the feedback process as part of their development practices. Given the potential scale of data, we will consider the potential to leverage a Large Language Model (LLM) to assist in dynamic prediction and summarization in addition to other emerging and established technologies. The impact of the model will be assessed by comparison to baseline metrics such as deployment frequency, defect escapes, production incidents, customer satisfaction and product adoption.

## V. RESEARCH PLAN

Given the active involvement of the researcher, we propose an Action Research methodology, with qualitative approaches used to explore high level themes and issues on the subject across Dell Technologies, and quantitative data to establish baseline metrics before defining a model for initial implementation. These data collection approaches will be further leveraged throughout the artifact iteration and refinement process. The researcher proposes to adopt a suitable Action Research approach, such as that of Agile Action Research [22]

### A. Exploratory Qualitative Phase

The first stage of data collection is a qualitative exploratory study within the organization. This approach will initially comprise of focus groups to identify broad themes and issues, with participation across a broad range of development and business stakeholders. Interviews with subject matter experts and key leaders will follow, with a view to refining the themes and issues identified in the focus groups. The primary objective of this exploratory phase is to inform design of the initial artifact for implementation.

### B. Baseline Quantitative Phase

Informed by the first phase of data collection, technical and systems data from sources such as *Jira, Gitlab, ServiceNow, Pendo, Dynatrace* and others will be leveraged in order to establish a baseline prior to implementation of a model/artifact.

*C. Artifact/Model Development, Test and Iteration*

Building on the qualitative and quantitative phases of data collection, an initial artifact will be developed and implemented in order to test the hypothesis. Both qualitative and quantitative approaches will be leveraged throughout the iteration process whilst continuously testing the hypothesis.

## VI. CONCLUSION

The research presented seeks to bridge the gap between the promises of agile software delivery and large-scale continuous software development. By identifying and addressing key challenges to customer validation and feedback integration in Dell Technologies, this work aims to enhance the agility, responsiveness, and competitiveness of software development processes in such contexts. I welcome the feedback of the symposium in guiding the future direction of the project.

## ACKNOWLEDGMENT

This doctoral research project is supported by Dell Technologies and SOLAS, an initiative of the Government of Ireland. I am extremely grateful for the continued support of my project mentors within Dell Technologies (Bob MacFarlane, Donal Golden) in addition to the invaluable support of my academic supervisors.